\documentclass[twocolumn,showpacs,preprintumbers,amsmath,amssymb,prb]{revtex4}
\usepackage{graphicx}
\usepackage{dcolumn}
\usepackage{bm}
\begin{document}

\title{Tunneling spectroscopy studies of aluminum oxide tunnel barrier layers}

\author{P. G. Mather}
\author{A. C. Perrella}
\author{E. Tan}
\author{J. C. Read}
\author{R. A. Buhrman}
\affiliation{%
School of Applied and Engineering Physics, Cornell University,
Ithaca NY 14853-2501
}%
\pacs{73.40.Gk, 73.61.Ng, 85.25.Cp, 68.37.Ef, 85.75.Dd}
\date{\today}

\begin{abstract}
We report scanning tunneling microscopy and ballistic electron
emission microscopy studies of the electronic states of the
uncovered and chemisorbed-oxygen covered surface of AlO$_x$ tunnel
barrier layers.  These states change when chemisorbed oxygen ions
are moved into the oxide by either flood gun electron bombardment
or by thermal annealing. The former, if sufficiently energetic,
results in locally well defined conduction band onsets at $\sim$ 1
V, while the latter results in a progressively higher local
conduction band onset, exceeding 2.3 V for 500 and 600 C thermal
anneals.
\end{abstract}

\maketitle

The prevalence of aluminum oxide layers formed by room temperature
oxidation as the barrier in Josephson junctions
(JJ)\cite{gurvitch} and magnetic tunnel junctions
(MTJs)\cite{moodera} continues to motivate efforts to better
understand and control its electronic structure. While bulk,
stoichiometric Al$_2$O$_3$ has a band gap of $\sim$ 8.8
eV,\cite{french} for amorphous AlO$_x$ films grown at $\sim$ 20 C
it is a much smaller.  This is beneficial as thin, transparent
barriers provide the high critical current densities (JJs) and low
specific impedance levels (MTJs) required by many applications,
but band tails, localized states, and spatial inhomogeneities that
may also be found in amorphous AlO$_x$\cite{plisch, rippard,
perrella} can be very detrimental for high performance, low noise
applications.\cite{zhang, tsymbal} Indeed, conducting atomic force
microscopy studies of AlO$_x$ layers have shown inhomogeneous
current distributions at the nanoscale, attributed to either a
variation in local barrier heights\cite{ando} or in barrier
thickness\cite{luo}. However, a serious challenge for such surface
spectroscopy studies of the electronic properties of AlO$_x$ is
that the surface is invariably covered, even in ultra-high vacuum
(UHV), with chemisorbed oxygen bound by positively charged oxygen
vacancies in the oxide, with the degree of coverage depending on
oxide thickness.\cite{perrella,eileen}

We report the use of scanning tunneling microscopy (STM) and
ballistic electron emission microscopy (BEEM) to examine the
density of states (DOS) of the AlO$_x$ surface, and to determine
how these states change when chemisorbed oxygen ions are moved
into the oxide by either flood-gun electron-bombardment (FGEB), or
by thermal annealing. Both treatments greatly reduce, if not
eliminate, low energy band tail states and narrow the DOS
distribution over an oxide area. However FGEB, which we argue has
similarities in effect to depositing a metallic over-layer with a
high work function $\phi$, causes different changes in the DOS
than annealing.  The former, if sufficiently energetic, results in
locally well defined conduction band onsets at $\sim$ 1 V, while
the latter results in a progressively higher local conduction band
onset, exceeding 2.3 V for 500 and 600 C anneals.

We fabricated the samples for this study via thin film thermal
evaporation and post-growth processing in UHV.  For most samples,
we deposited 12 nm of Au on hydrogen terminated (111) Si to form a
high quality Schottky barrier (SB) to serve as the BEEM detector.
This was followed by a 1.2 nm buffer layer of Cu, 1.2 nm Co, and
finally 1 nm Al, which was oxidized by a 10 torr-sec exposure to
oxygen (99.9985 \% purity).  X-ray photoemission (XPS)
measurements show that this exposure forms a $\sim$ 1 nm AlOx
layer. Some samples were then annealed in UHV while others were
subjected to FGEB.  Upon completion of processing, the sample was
vacuum transferred to an adjacent UHV chamber for STM and BEEM
measurements.  Samples that were annealed at T$_a \geq$ 500 C had
the Au/Cu/Co underlayers replaced with a single Co 20 nm layer to
avoid Au diffusion into the Si.  No BEEM measurements were made on
those samples.

Scanning tunneling spectroscopy (STS) studies of aluminum oxide
barrier layers are challenging because of both the chemisorbed
oxygen discussed above and the propensity of the tunneling current
to locally degrade the oxide via hot electron effects. It was
shown earlier\cite{perrella} that BEEM measurements, in
conjunction with STS, can distinguish between where the STM is
tunneling to the chemisorbed oxygen that is stabilized into
nanoscale clusters by a local maximum in the vacancy
concentration, and where the STM is tunneling to the bare oxide
surface.  In the former case, at sufficiently high tip bias, the
electrons tunnel predominately into oxygen-cluster states, from
which they eventually exit to the underlying metal via an
inelastic process, resulting in no discernable BEEM signal. In the
latter case, due to a low density of extended states on the oxide,
the STM tip must approach close enough to establish feedback that
there is substantial tunnel current directly through the oxide
layer to the base electrode. Thus, a fraction of the
tunnel-injected electrons travel ballistically to and across the
underlying SB if they have the requisite energy and momentum
values.  The result for untreated, oxidized samples is BEEM
spectra that are the same in form but reduced in amplitude from
those obtained from unoxidized samples.

\begin{figure}
\includegraphics[width=8.5cm]{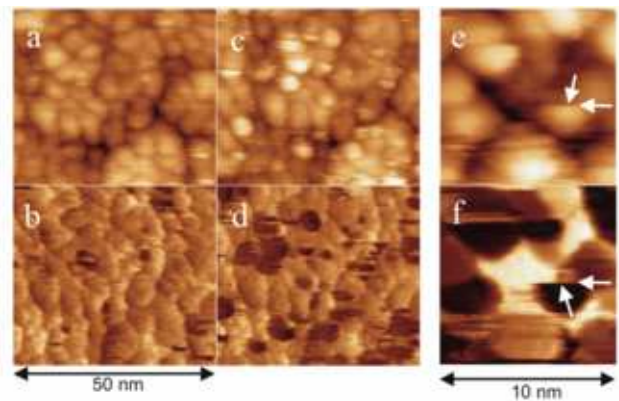}
\caption{(Color online) STM topographic images (top)  and BEEM
current images (bottom) for a Si (111)/Au (12nm)/Cu (1.2nm)/Co
(1.2nm)/Al (1nm) sample exposed to a 10 torr-sec high purity
oxygen dose.  (a) and (b)  is the initial scan while (c) and (d)
are a later scan.  The dark regions in BEEM images are clusters of
chemisorbed O$_2^-$  arising from STM stress-induced vacancy
generation. (e) and (f) show images for a vacuum-annealed sample
(3 min. at 375 C).  The abrupt vertical changes in the images (see
e.g. arrows in high resolution image) are due to clusters of
chemisorbed oxygen suddenly moving under the tip as the scan
proceeds}
\end{figure}
BEEM is also effective in detecting degradation of the oxide layer
due to electrical stress during STM studies. The primary effect of
prolonged and/or high current STM measurement over a small region
of the surface is to partially reduce the oxide, generating
positively charged oxygen vacancies, which can locally stabilize a
nanocluster of chemisorbed O$_2^-$ on the surface. This is readily
detected by BEEM if the surface is scanned at high enough bias
that the tunneling is to unoccupied states. Fig. 1 (a) and (b) are
initial topographic and BEEM images taken on the surface of an
un-annealed sample and show a low density of chemisorbed clusters;
dark regions in the BEEM scan correlated with elevated regions in
the topography. Subsequent scans, e.g. Fig. 1 (c) and (d), reveal
a stress-induced accumulation of such clusters.  These stress
effects which can cause the motion of a cluster to a more stable
position as oxygen vacancies are generated under the tip, require
that STS on the uncovered surface be taken only for a short time
and with a low tunneling current, impacting both the STS
signal-to-noise and the ability to systematically survey the oxide
surface.  Fig. 1 (e) and (f) show images of a thermally annealed
sample (3 min at 375 C).  While the un-annealed sample accumulates
an increasing number of clusters with scanning time, annealed
samples exhibit only the motion of an approximately constant
density of clusters, demonstrating that the annealed oxide is
resistant to STM stress effects.

Using BEEM to determine locations where the tunneling is to the
uncovered oxide surface, STS and BEEM measurements were made on
the as grown oxide, and after different surface processing
procedures.  Fig. 2 shows typical results of the differential
logarithmic conductivity, $d\ln(I_t)/d\ln(V_t)$, proportional to
the local DOS of the surface, together with the BEEM spectrum,
$I_{c}(V_{t})$. The tip was held steady during measurement, with
$\sim$ 1 nm lateral drift, and 20 $I_{t}(V_{t})$ scans were
averaged. As illustrated in Fig. 2a, the uncovered, as-formed
AlO$_x$ surface yields DOS curves with a nearly parabolic
dependence on bias extending to within 150 mV of the Fermi level
E$_f(V_t =0)$. While the details of the local DOS varied across
the uncovered oxide, at no location was the DOS onset higher than
200 mV.

However, the STS is not indicative of the electronic structure of
the oxide when it is embedded in a tunnel junction structure via
over-coating with an electrode. An XPS study\cite{eileen} has
found that this drives much of the chemisorbed oxygen into the
oxide, filling oxygen vacancy sites and presumably substantially
altering the electronic character of the tunnel barrier layer.  A
strong, positive correlation was found between the increase in
oxygen content and the work function difference between top and
bottom electrodes.  This is consistent with the finding that
negatively charging the surface via FGEB can also drive some or
all of the chemisorbed oxygen into the oxide, depending upon the
bias and exposure time.

To determine the effect of filling oxide vacancies on the
electronic structure of AlO$_x$ layers, samples were subjected to
FGEB and then examined by STS and BEEM.  Samples were subjected to
either 10 V or 20 V FGEB at a current density of $\sim$ 20
$\mu$/cm$^2$ for 2 hours. The 10 eV sample was dosed again (300
torr-sec), and bombarded a second time, which further stabilizes
oxygen in the AlO$_x$ barrier.\cite{eileen}
\begin{figure}
\includegraphics[width=8.5cm]{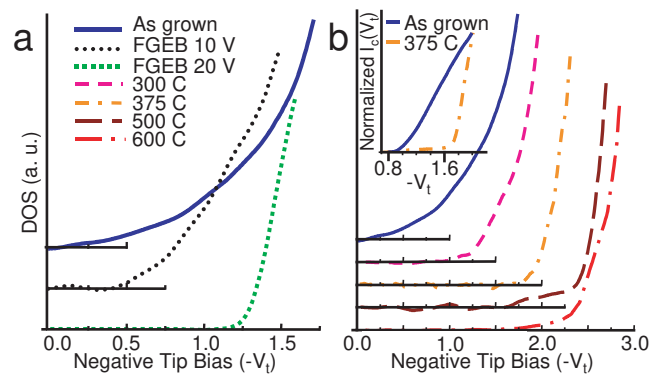}
\caption{(Color online) Differential logarithmic conductivity,
$d\ln(I_t)/d\ln(V_t)$ acquired from STS (a and b) and BEEM
$I_{c}(V_{t})$ (inset) for thermally oxidized aluminum samples
with different anneal temperature or flood gun electron
bombardment (FGEB) voltages.  As the bombardment voltage
increases, more chemisorbed O$_2^-$ clusters are driven into
oxygen vacancies in the aluminum oxide resulting in a better
defined conduction band edge. Annealed samples exhibit a larger
oxide band gap as anneal temperature increases.}
\end{figure}
As illustrated in Fig. 2a, the effect of FGEB on the electronic
properties of the oxide was profound.  After the 10 V treatment
STS measurements on the uncovered oxide surface showed no
detectable oxide states from 0 (E$_f$) to $\sim$ 0.5 V, with a
roughly parabolic energy dependence above 0.5 V. The details of
the DOS varied slightly with location, but at no point was the
onset either lower than 0.35 V or higher than 0.65 V. However, due
to the inability of the STS measurements to completely survey the
oxide surface and to the limited sensitivity of the measurement,
we cannot rule out a small density of remnant low energy states in
the oxide. STS measurements made with both positive and negative
tip bias showed that the local DOS is roughly symmetric about V =
0, indicating that the Fermi level is pinned near the center of
the oxide band-gap for these AlO$_x$ layers.

For the 20 V FGEB the region of zero DOS increased further to
approximately 1.1 V, above which the DOS increased rapidly, as
shown in Fig. 2a.  A notable feature of the 20 V sample was that
it was particularly quick to degrade; forming chemisorbed oxygen
clusters shortly after we began examining a region with STM,
indicating that when oxygen has been forcefully driven into the
oxide it is unstable against electronic stress.  At no point did a
DOS measurement on a fresh region of the oxide show an onset lower
than 0.8 V, or higher than 1.2 V.

We conclude that the low energy oxide states are strongly
correlated with the vacancy concentration in the oxide; as these
positively charged sites are progressively filled with oxygen ions
the band gap grows.  What is somewhat surprising is that the band
gap is relatively uniform from spot to spot as examined by STS,
both for the 10 V FGEB sample, which only partially removed the
chemisorbed oxygen, and for the 20 V FGEB sample. This indicates
that a relatively uniform driving force is responsible for the
formation of the vacancies in the oxide as it grows, and that the
electric field established across the oxide by the FGEB is rather
uniformly effective in filling these vacancies.

Thermal annealing of AlO$_x$ layers had quantitatively different
effects.  XPS shows that the chemisorbed oxygen and metallic
aluminum signals decreased while the oxidic aluminum  signal
increased with increasing anneal temperature T$_a$, but some
chemisorbed oxygen still remained after a 3 minute anneal at 500
C, indicating that while the oxide grows thicker, it still had
positive charged vacancies at that point.  Typical results from
STS measurements on uncovered regions of the oxide surface after 3
minute anneals at different T$_a$ are shown in Fig. 2b.  The
effect of increasing T$_a$ was to progressively increase the DOS
threshold, from $\sim$ 1.2 V for T$_a$ = 300 C, to $\sim$ 1.8 V
for T$_a$ = 375 C. Further insight into the electronic structure
of the 375 C sample is gained through examination of the BEEM
$I_c(V_t)$ (inset), where, in addition to the SB threshold at
$\sim$ 0.8 V, a second threshold is observed at $\sim$ 1.7 V,
suggesting the onset of extended states that allow ballistic
transport through the oxide layer above that bias. The presence of
localized states on the oxide surface has been substantially
reduced, leaving mostly extended states through which ballistic
transport through the oxide can take place. However, for T$_a$ =
500 and 600 C, while the main DOS threshold moved up to $\sim$
2.25 V, in both these cases there was generally a lower DOS
"band-tail" that extended approximately 0.5 V below this main
threshold point, indicating the presence of some residual defect
states or disorder in the oxide, even at these relatively high
anneal temperatures.

Valence band XPS and transmission electron microscopy studies of
$\sim$ 4 nm, AlO$_x$ layers have found a transition from an
amorphous form for a 100 C growth temperature to largely
$\gamma$-alumina for 400 C growth.\cite{snijders} The large
increase in the oxide band gap observed when T$_a$ is increased to
375 C and above is in accordance with those studies if we
attribute the DOS difference between the 20 V FGEB sample and the
high temperature annealed samples as arising from the former being
amorphous and the latter largely crystalline.  There is also a
pronounced increase in the stability of the annealed oxide layers;
these oxides are much slower to develop chemisorbed clusters while
being studied by STM.

While FGEB and annealing the AlO$_x$ barrier layer prior to
deposition of the top electrode are not standard procedures in
tunnel junction formation, the use of top electrodes with a higher
$\phi$ than the metal immediately under the oxide layer, and of
mild annealing (= 400 C) after top electrode deposition are common
approaches for JJs and MTJs respectively. The STS results reported
here indicate what effects these approaches have on the electronic
properties of the barrier. The use of a top electrode with a
higher $\phi$, drives more chemisorbed oxygen into the oxide,
filling vacancy sites.\cite{eileen} This removes low energy states
from the oxide and increases the tunnel barrier height. Mild
annealing of tunnel barriers with symmetric electrodes can have a
similar effect by more uniformly distributing chemisorbed oxygen
that the deposition of the metallic over-layer may trap at and
near the top of the oxide layer.\cite{sousa}  Thus, electrode
oxidation, caused by over-oxidation and reaction with the
chemisorbed layer in favor of enhanced oxidation of the AlO$_x$,
is reduced. This also removes low energy states from the oxide and
raises the barrier height. Thermal annealing also lowers atomic
disorder in the barrier. While BEEM studies of buried layers
indicate that the barrier layer is spatially rather uniform after
the deposition of the top electrode onto the chemisorbed oxygen
covered oxide surface, it may prove advantageous to process the
oxide layer by FGEB or thermal annealing prior to this deposition
step.  Indeed a preliminary study has already found that as much
as a factor of ten reduction of 1/f resistance noise amplitude in
Al/AlO$_x$/Al junctions can be effected by FGEB of the oxide
layer.\cite{eileen}

This research was supported by the Office of Naval Research, the
ARO/MURI program, and by DARPA/DSO.  The research benefited from
use of the facilities of the Center for Nanoscale Systems,
supported by NSF through the NSEC program, and of the Cornell
Nanoscale Facility/NNIN, also supported by NSF.

\bibliography{annealed}

\begin{thebibliography}{13}
\expandafter\ifx\csname natexlab\endcsname\relax\def\natexlab#1{#1}\fi
\expandafter\ifx\csname bibnamefont\endcsname\relax
  \def\bibnamefont#1{#1}\fi
\expandafter\ifx\csname bibfnamefont\endcsname\relax
  \def\bibfnamefont#1{#1}\fi
\expandafter\ifx\csname citenamefont\endcsname\relax
  \def\citenamefont#1{#1}\fi
\expandafter\ifx\csname url\endcsname\relax
  \def\url#1{\texttt{#1}}\fi
\expandafter\ifx\csname urlprefix\endcsname\relax\def\urlprefix{URL }\fi
\providecommand{\bibinfo}[2]{#2}
\providecommand{\eprint}[2][]{\url{#2}}

\bibitem[{\citenamefont{Gurvitch et~al.}(1983)\citenamefont{Gurvitch,
  Washington, and Huggins}}]{gurvitch}
\bibinfo{author}{\bibfnamefont{M.}~\bibnamefont{Gurvitch}},
  \bibinfo{author}{\bibfnamefont{M.~A.} \bibnamefont{Washington}},
  \bibnamefont{and} \bibinfo{author}{\bibfnamefont{H.}~\bibnamefont{Huggins}},
  \bibinfo{journal}{Appl.\ Phys.\ Lett.} \textbf{\bibinfo{volume}{42}},
  \bibinfo{pages}{472} (\bibinfo{year}{1983}).

\bibitem[{\citenamefont{Moodera et~al.}(1995)\citenamefont{Moodera, Kinder,
  Wong, and Meservey}}]{moodera}
\bibinfo{author}{\bibfnamefont{J.~S.} \bibnamefont{Moodera}},
  \bibinfo{author}{\bibfnamefont{L.}~\bibnamefont{Kinder}},
  \bibinfo{author}{\bibfnamefont{T.}~\bibnamefont{Wong}}, \bibnamefont{and}
  \bibinfo{author}{\bibfnamefont{R.}~\bibnamefont{Meservey}},
  \bibinfo{journal}{Phys.\ Rev.\ Lett.} \textbf{\bibinfo{volume}{74}},
  \bibinfo{pages}{3273} (\bibinfo{year}{1995}).

\bibitem[{\citenamefont{French}(1990)}]{french}
\bibinfo{author}{\bibfnamefont{R.~H.} \bibnamefont{French}},
  \bibinfo{journal}{J.\ Am.\ Ceram.\ Soc.} \textbf{\bibinfo{volume}{73}},
  \bibinfo{pages}{477} (\bibinfo{year}{1990}).

\bibitem[{\citenamefont{Plisch et~al.}(2001)\citenamefont{Plisch, Chang,
  Silcox, and Buhrman}}]{plisch}
\bibinfo{author}{\bibfnamefont{M.~J.} \bibnamefont{Plisch}},
  \bibinfo{author}{\bibfnamefont{J.~L.} \bibnamefont{Chang}},
  \bibinfo{author}{\bibfnamefont{J.}~\bibnamefont{Silcox}}, \bibnamefont{and}
  \bibinfo{author}{\bibfnamefont{R.~A.} \bibnamefont{Buhrman}},
  \bibinfo{journal}{Appl.\ Phys.\ Lett.} \textbf{\bibinfo{volume}{79}},
  \bibinfo{pages}{391} (\bibinfo{year}{2001}).

\bibitem[{\citenamefont{Rippard et~al.}(2002)\citenamefont{Rippard, Perrella,
  Albert, and Buhrman}}]{rippard}
\bibinfo{author}{\bibfnamefont{W.~H.} \bibnamefont{Rippard}},
  \bibinfo{author}{\bibfnamefont{A.~C.} \bibnamefont{Perrella}},
  \bibinfo{author}{\bibfnamefont{F.~J.} \bibnamefont{Albert}},
  \bibnamefont{and} \bibinfo{author}{\bibfnamefont{R.~A.}
  \bibnamefont{Buhrman}}, \bibinfo{journal}{Phys.\ Rev. \ Lett.}
  \textbf{\bibinfo{volume}{88}}, \bibinfo{pages}{046805}
  (\bibinfo{year}{2002}).

\bibitem[{\citenamefont{Perrella et~al.}(2002)\citenamefont{Perrella, Rippard,
  Mather, Plisch, and Buhrman}}]{perrella}
\bibinfo{author}{\bibfnamefont{A.~C.} \bibnamefont{Perrella}},
  \bibinfo{author}{\bibfnamefont{W.~H.} \bibnamefont{Rippard}},
  \bibinfo{author}{\bibfnamefont{P.~G.} \bibnamefont{Mather}},
  \bibinfo{author}{\bibfnamefont{M.~J.} \bibnamefont{Plisch}},
  \bibnamefont{and} \bibinfo{author}{\bibfnamefont{R.~A.}
  \bibnamefont{Buhrman}}, \bibinfo{journal}{Phys.\ Rev.\ B.}
  \textbf{\bibinfo{volume}{675}}, \bibinfo{pages}{201403}
  (\bibinfo{year}{2002}).

\bibitem[{\citenamefont{Zhang and White}(1998)}]{zhang}
\bibinfo{author}{\bibfnamefont{J.}~\bibnamefont{Zhang}} \bibnamefont{and}
  \bibinfo{author}{\bibfnamefont{R.~M.} \bibnamefont{White}},
  \bibinfo{journal}{J.\ Appl.\ Phys.} \textbf{\bibinfo{volume}{83}},
  \bibinfo{pages}{6512} (\bibinfo{year}{1998}).

\bibitem[{\citenamefont{Tsymbal and Pettifor}(1998)}]{tsymbal}
\bibinfo{author}{\bibfnamefont{E.}~\bibnamefont{Tsymbal}} \bibnamefont{and}
  \bibinfo{author}{\bibfnamefont{D.}~\bibnamefont{Pettifor}},
  \bibinfo{journal}{Phys.\ Rev.\ B.} \textbf{\bibinfo{volume}{58}},
  \bibinfo{pages}{2533} (\bibinfo{year}{1998}).

\bibitem[{\citenamefont{Ando et~al.}(2000)\citenamefont{Ando, Kubota, Hayashi,
  Kamijo, Yaoita, Yu, Han, and Miyazaki}}]{ando}
\bibinfo{author}{\bibfnamefont{Y.}~\bibnamefont{Ando}},
  \bibinfo{author}{\bibfnamefont{H.}~\bibnamefont{Kubota}},
  \bibinfo{author}{\bibfnamefont{M.}~\bibnamefont{Hayashi}},
  \bibinfo{author}{\bibfnamefont{M.}~\bibnamefont{Kamijo}},
  \bibinfo{author}{\bibfnamefont{K.}~\bibnamefont{Yaoita}},
  \bibinfo{author}{\bibfnamefont{A.~C.~C.} \bibnamefont{Yu}},
  \bibinfo{author}{\bibfnamefont{X.-F.} \bibnamefont{Han}}, \bibnamefont{and}
  \bibinfo{author}{\bibfnamefont{T.}~\bibnamefont{Miyazaki}},
  \bibinfo{journal}{Jpn.\ J.\ Appl.\ Phys.} \textbf{\bibinfo{volume}{39}},
  \bibinfo{pages}{5832} (\bibinfo{year}{2000}).

\bibitem[{\citenamefont{Luo et~al.}(2001)\citenamefont{Luo, Wong, Pakhomov, Xu,
  Wilson, and Wong}}]{luo}
\bibinfo{author}{\bibfnamefont{E.~Z.} \bibnamefont{Luo}},
  \bibinfo{author}{\bibfnamefont{S.~K.} \bibnamefont{Wong}},
  \bibinfo{author}{\bibfnamefont{A.~B.} \bibnamefont{Pakhomov}},
  \bibinfo{author}{\bibfnamefont{J.~B.} \bibnamefont{Xu}},
  \bibinfo{author}{\bibfnamefont{I.~H.} \bibnamefont{Wilson}},
  \bibnamefont{and} \bibinfo{author}{\bibfnamefont{C.~Y.} \bibnamefont{Wong}},
  \bibinfo{journal}{J.\ Appl.\ Phys.} \textbf{\bibinfo{volume}{90}},
  \bibinfo{pages}{5202} (\bibinfo{year}{2001}).

\bibitem[{\citenamefont{Tan et~al.}(2005)\citenamefont{Tan, Mather, Perrella,
  Read, and Buhrman}}]{eileen}
\bibinfo{author}{\bibfnamefont{E.}~\bibnamefont{Tan}},
  \bibinfo{author}{\bibfnamefont{P.~G.} \bibnamefont{Mather}},
  \bibinfo{author}{\bibfnamefont{A.~C.} \bibnamefont{Perrella}},
  \bibinfo{author}{\bibfnamefont{J.~C.} \bibnamefont{Read}}, \bibnamefont{and}
  \bibinfo{author}{\bibfnamefont{R.~A.} \bibnamefont{Buhrman}},
  \bibinfo{journal}{cond-mat/0501354}  (\bibinfo{year}{2005}).

\bibitem[{\citenamefont{Snijders et~al.}(2002)\citenamefont{Snijders, Jeurgens,
  and Sloof}}]{snijders}
\bibinfo{author}{\bibfnamefont{P.~C.} \bibnamefont{Snijders}},
  \bibinfo{author}{\bibfnamefont{L.~P.~H.} \bibnamefont{Jeurgens}},
  \bibnamefont{and} \bibinfo{author}{\bibfnamefont{W.~G.} \bibnamefont{Sloof}},
  \bibinfo{journal}{Surface\ Science} \textbf{\bibinfo{volume}{496}},
  \bibinfo{pages}{97} (\bibinfo{year}{2002}).

\bibitem[{\citenamefont{Sousa et~al.}(1998)\citenamefont{Sousa, Sun, Soares,
  Freitas, Kling, da~Silva, and Soares}}]{sousa}
\bibinfo{author}{\bibfnamefont{R.~C.} \bibnamefont{Sousa}},
  \bibinfo{author}{\bibfnamefont{J.~J.} \bibnamefont{Sun}},
  \bibinfo{author}{\bibfnamefont{V.}~\bibnamefont{Soares}},
  \bibinfo{author}{\bibfnamefont{P.~P.} \bibnamefont{Freitas}},
  \bibinfo{author}{\bibfnamefont{A.}~\bibnamefont{Kling}},
  \bibinfo{author}{\bibfnamefont{M.~F.} \bibnamefont{da~Silva}},
  \bibnamefont{and} \bibinfo{author}{\bibfnamefont{J.~C.}
  \bibnamefont{Soares}}, \bibinfo{journal}{Phys.\ Rev.\ B.}
  \textbf{\bibinfo{volume}{73}}, \bibinfo{pages}{3288} (\bibinfo{year}{1998}).

\end{thebibliography}
\end{document}